\documentstyle[12pt,epsfig]{article} 
\def \LSP{\widetilde{N}_1}
\def \N2{\widetilde{N}_2}

\def \W1{\widetilde{\chi}_1^{\pm}}
\def \WC2{\widetilde{\chi}_2^{\pm}}
\def \SNU{\tilde{\nu}}

\def \SQ{\tilde{q}}
\def \LSQ{\tilde{q}_L}
\def \DR{\tilde{d}_R}

\def \GL{\tilde{g}}

\def \MN2{M_{\widetilde{N}_2}}
\def \MCH1{M_{\widetilde{\chi}_1^{\pm}}}
\def \M2CH{M_{\widetilde{\chi}_2^{\pm}}}

\def \MSNU{m_{\tilde{\nu}}}
\def \MSELR{m_{\tilde{e}_R}}

\def \MLL{m_{\tilde{l}_L}}
\def \MLR{m_{\tilde{l}_R}}
\def \GLUM{m_{\tilde{g}}}

\def \MSQ{m_{\tilde{q}}}
\def \LMSQ{m_{\tilde{q_L}}}
\def \MDL{m_{\tilde{d}_L}}
\def \MUL{m_{\tilde{u}_L}}
\def \MDR{m_{\tilde{d}_R}}
\def \MUR{m_{\tilde{u}_R}}

\def \MS2{m_{\tilde{q}}^2}
\def \MSUP2{m_{\tilde{u}}^2}
\def \MSDN2{m_{\tilde{d}}^2}
\def \MSEL2{m_{\tilde{e}}^2}
\def \MSLEP2{m_{\tilde{l}}^2}

\def \SUSY{\emph{SUSY }}
\def \SUGRA{\emph{SUGRA }}
\def \SM{\emph{SM }}

\begin{document}

\begin{flushright}
IC/97/207\\ 
hep-ph/yyyyyyy
\end{flushright}

\vskip 50pt
\begin{center}
{\bf Signatures of Non-Universal Soft Breaking Sfermion Masses \\
at Hadron Colliders}
\end{center}

\vskip 35pt
\begin{center}
Amitava Datta \footnote{E-mail: adatta@juphys.ernet.in} and 
Aseshkrishna Datta \footnote{E-mail: asesh@juphys.ernet.in}\\
Department of Physics, Jadavpur University \\
Calcutta- 700 032, India,\\
\vskip 15pt
and
\vskip 15pt
M.K. Parida \footnote{E-mail:mparida@nehus.ren.nic.in} \footnote{Permanent
Address:Physics Department,North-Eastern Hill University,Shillong 793022,India}
\footnote{ICTP Associate}\\
International Centre for Theoretical Physics, Trieste, Italy.
\end{center}

\vskip 30pt

\begin{abstract}
We identify several mass patterns, within the framework of $N=1$
\SUGRA  with nonuniversal soft breaking masses for the sfermions, which may
significantly alter \SUSY  signals and the current squark-gluino mass
limits from the Tevatron. These effects are illustrated in a $SO(10)$
\SUSY  GUT with an intermediate mass scale, but the conclusions are also valid
in \SUSY  $SO(10)$ models with grand deserts.
\end{abstract}

\newpage
A strong prediction of $N$=1 supergravity (\SUGRA) models with
supersymmetry (\SUSY) breaking due to the hidden sector is that the soft
breaking parameters are universal \cite{R1}. The economy in the number of
parameters and the resulting predictive power have made this model
(referred to hereafter as the conventional scenario) particularly popular.
Even the current experimental limits on the sparticle masses are often
derived by assuming that their spectrum is indeed given by the
conventional scenario.

It has, however, been emphasised in the literature  for quite some time that
non-universal soft \SUSY  breaking terms at some high scale may arise quite
naturally within the \SUGRA framework \cite{R2}.
For example, even if strict universality holds at the Planck
scale,  non-universal soft breaking terms may arise at a lower scale 
due to renormalization effects. Within the frame work of a grand unified
theory (GUT) such effects are calculable. One simply has to consider
the evolution from the Planck scale to the GUT scale ($M_G$) by using the 
Renormalisation Group (RG) equations of the underlying GUT \cite{R3}. It is, 
therefore,
extremely important to study the impact of this nonuniversality on collider 
signatures of sparticle production.

Once we accept the possibility of  non-universality of the soft breaking
parameters of the full theory including heavy fields, through the above or
any other suitable mechanism, an interesting consequence follows whenever
the rank of the GUT group and/or some intermediate symmetry is reduced by
spontaneous symmetry breaking \cite{R4,R5}. One obtains  $D$-term
contributions, which apriori can be quite significant in magnitude,
leading to non-universal squark and slepton masses at a scale much higher
than the electroweak(EW) scale. As a result the low energy sparticle spectrum
can be significantly different from that in the conventional scenario.

Some phenomenological consequences of these $D$-terms in the context of \SUSY 
GUTs with grand deserts as well as models with intermediate mass scales have
been considered in the literature \cite{R5}. In this letter,
 we focus on those aspects of the sparticle spectrum which may lead to 
{\bf collider signatures significantly  distinct from the conventional ones}
 and, hence,  {\bf to revisions of the current mass limits of the sparticles}.

For the purpose of illustration we have carried out our calculations in a
$SO(10)$ \SUSY  GUT model with an intermediate mass scale ($M_I$)\cite{R6}
which utilises the particle spectrum predicted from superstring
compactification\cite{R7}. It should , however, be emphasised that our
conclusions are not crucially dependent on the intermediate scale.  As we
shall discuss below, many of our conclusions hold even if the GUT group
breaks down to the standard model (\SM) gauge group through a single step.

The model with an intermediate scale that we have considered is highly
attractive for various aesthetic and phenomenological reasons. It provides
a natural explanation for small neutrino masses through the see-saw
mechanism \cite{R8}, which has been shown to operate in string-inspired
$SO(10)$ in the absence of the higher dimensional representation 126
\cite{R9}. With $M_I\simeq 10^{12}-10^{13}$ GeV, the predicted neutrino
mass spectrum  allows ${\nu}_{\tau}$ as a prospective candidate for hot
dark matter of the universe and an explanation of the solar neutrino
puzzle through ${\nu}_e-{\nu}_{\mu}$ oscillation in a manner similar to
that discussed at length for intermediate scale models \cite{R10} via MSW
mechanism \cite{R11}.

Among the various candidates for the intermediate gauge group the Pati -
Salam group $G_{224}=SU(2)_L \otimes SU(2)_R \otimes SU(4)_c$ with
$g_{2L}=g_{2R}$
\cite{R12} has many appealing features. It contains the seeds of
quark-lepton unification and involves only two gauge coupling constants
($g_{2L}=g_{2R}$, $g_{4C}$) which can be determined using low-energy data, 
most notably the ones from LEP. The boundary condition on Yukawa couplings 
necessary for the implementation of the see-saw mechanism holds starting from 
the GUT scale down to the intermediate scale.
Another important consequence of this model is that all major sources of
uncertainities like threshold  and gravitational effects as well as those due 
to radiative corrections from higher scale ($\mu > M_I$) do not affect the
predicted values of $sin^2 \theta_W$ \cite{R13} or $M_I$ \cite{R6} which 
have been proved through three theorems\cite{R6,R13}.

There are claims in the literature \cite{R14} that  the $G_{224}$ 
intermediate gauge symmetry with $M_I$ significantly different from $M_G$ 
is ruled out. Such an analysis utilises the right-handed triplet,
[$\Delta_R(1,3,\overline{10})]$, under $G_{224}$ , contained in the 
representation 126 of $SO(10)$ and its chiral conjugate, to break the 
intermediate gauge symmetry and to
implement the see-saw mechanism. The large contributions of $\Delta_R$ and
$\overline{\Delta_R}$ to the $\beta$ function disallow the
possibility of $M_I$ being substantially lighter than the unification
scale ($M_{G}$) \cite{R14,R15}. Further, it is well known that the presence of 
$126+\overline {\rm 126}$ spoils perturbative grand unification above $\mu \ge 
8M_G$ in \SUSY  $SO(10)$ and attempts of unification have been made
through superstring-inspired particle spectrum \cite{R9}.

Spontaneous breakdown of the intermediate gauge symmetry $G_{224}$ has also 
been carried out utilising the predicted representations \cite{R7} containing
${\rm16}+\overline{\rm{16}}$ of $SO(10)$. In this scenario the mass of the 
right handed majorana neutrino is generated either through
non-renormalizable $5$-dimensional opeartors  or through purely
renormalizable interaction by a novel mechanism \cite{R9}. The smaller
contribution of the left and the right-handed doublets to the one-loop $\beta$
function in this case permits grand unification at the string- scale 
$M_{G}\simeq
M_{string} \simeq 5\times 10^{17} \;GeV$, but having the intermediate scale
substantially different; $M_I \simeq 10^{12}- 10^{13} \; GeV$. Such an 
intermediate symmetry  needs the presence of lighter scalar superfields,
$\sigma^{\pm}_{R} (1,\pm 1,1),\chi_{3} (1,\frac{2}{3},3),
\overline{\chi}_{3}(1,-\frac{2}{3},\overline{3})$ under the standard gauge
group near the TeV scale and the $G_{224}$
submultiplet $\sigma_C(1,1,15)$ near the intermediate scale. While 
${\sigma}^{\pm}_R$
and  ${\sigma}_C \subset 45$ of $SO(10)$ ,${\chi}_3$ and $\overline {\chi}_3$
correspond to the would-be goldstone bosons 
$\subset {\rm 16}+\overline {\rm 16}$, 
not absorbed in the process of intermediate breaking.   

The contributions to the one loop $\beta$ functions  
in such a model at different energy scales turn
out to be 
\begin{equation}
M_Z< \mu< 1 \; TeV \;\;\;\;\;\;\;  b_i=\left( \begin{array}{c} 33/5 \\ 1 \\ -3 
\end{array} \right) 
\end{equation}

\begin{equation}
1 \; TeV< \mu< M_I= 10^{13} \; GeV  \;\;\;\;\;\;\;  b_i=\left( \begin{array}{c}
47/5 \\ 1 \\ -2 \end{array} \right) 
\end{equation}
where $i=Y$, $SU(2)_L$, $SU(3)_c$ and the extra light Higgs scalars, assumed to
have masses $\cal O$ (1 TeV), contributes in eq. (2), and

\begin{equation}
M_I< \mu< M_{G}  \;\;\;\;\;\;\;\;   b_i= \left( \begin{array}{c} 
7 \\ 7 \\ 2 \end{array} \right) 
\end{equation}

where $i=SU(2)_L \, , \, SU(2)_R \, , \, SU(4)_c$ and the discrete $L-R$
symmetry is assumed for $\mu > M_I$.

Using $sin^2\theta_W=0.2315 \pm 0.0003$, $\alpha^{-1}(m_Z)=128.9\pm 0.09$
and $\alpha^{-1}_{3c}(m_Z)=0.119\pm 0.004$ \cite{R16} and the above
$\beta$ functions in eqs (1-3) we find
$\alpha_{2R}(M_I)=0.039=\alpha_{2L}(M_I)$, $\alpha_{4c}(M_I)=0.059$ and
$\alpha_{GUT}=0.074 \pm  0.004$..

Using the gluino mass($\GLUM$) at the EW scale as an input and
the appropriate RG equations, we calculate the universal gaugino mass
$M_{1/2}$ at the GUT scale. We then evolve downwards to calculate $U(1)_Y$
and $SU(2)_L$ gaugino masses $M_1$ and $M_2$ at the EW scale.

In this calculation we have taken into account the evolutions $M_{G}
\to M_I$, $M_I \to TeV$ scale and the $TeV \; {\rm scale} \to M_Z$ with
appropriate $\beta$ functions given above. As shown  in eqs 
(4.2)--(4.6), of \cite{R5} new $D$-term contributions, involving only one
unknown parameter, introduce nonuniversality in the sfermion masses at $M_I$
for the first two generations and we focus our attention on the spectrum of 
this sector. In contrast the third generation and the Higgs scalars can acquire, in principle,
both $D$ and $F$-term contributions to their masses at $M_I$ \cite{R5}. 
These sectors involving more unknown parameters are not included in our analysis.
We shall, however, assume that $\tilde{b}_L$ and $\tilde{b}_R$ are
approximately degenerate with $\tilde{d}_{L,R}$
and their masses at $M_I$ are also given  by the formulae given below.
 We need this simplifying assumption to compare with the
conventional phenomenology which assumes 5 degenerate squark flavours of both 
$L$ and $R$ type at a high scale. For ${\mu}\ge M_I$, the assumption 
of left-right discrete symmetry in $G_{224}$ constrains the masses of the left
and the right chiral multiplets to be equal($m_L=m_R=m_0$), and, as compared 
to the formula given in \cite{R5}, the number of parameters is reduced
by one. 

Using the values of the gauge-coupling constants given above we find 
for the masses of one generation of sfermions \cite{R5} at $\mu=M_I$: 
\begin{eqnarray}
\MS2 &=& m_0^2+0.74 D  \nonumber \\
\MSUP2 &=& m_0^2 + 0.24 D  \nonumber \\
\MSEL2 &=& m_0^2 + 1.24 D  \nonumber \\
\MSLEP2 &=& m_0^2 - 2.22 D  \nonumber \\
\MSDN2 &=& m_0^2 - 1.72 D 
\end{eqnarray}
where $\SQ, \tilde{l} $ stand for doublets of squarks and sleptons, while
$\tilde{d}, \tilde{u} $ and $\tilde{e}$ denote right-handed 
singlet squarks and sleptons.

In the above eqs. $D$ is an unknown parameter which may have either sign.
The requirement of no tachyonic degrees of freedom at $M_I$, however,
implies $D< 0.45 {m_0}^2$ (for $D >0$) from $\MSLEP2$ and $D >- 0.81 {m_0}^2$
(for $D< 0$) from $\MSEL2$. In the subsequent phenomenological analyses, 
values of $D$ between these two extremes will be considered.

 From eqs. (4) it follows that for $ D > 0 $, the lepton doublet can be
considerably lighter than the rest of the sfermions at $M_I$. Another
notable feature is the suppression of $\MSDN2$ compared to  other
squark masses. Suppression also affects $\MSUP2$ although its magnitude
 is rather modest. On the other hand $D< 0 $, leaves open
the possibility of a relatively light $\tilde{e}$ while the $\SQ$' s become
the lightest squarks. We note that most of these features also
hold at $M_G$ qualitatively if additional $D$-term contributions are 
taken into account
in the one step breaking of \SUSY  $SO(10)$ into the \SM  gauge group (see
eqs (4.9) - (4.13) of \cite{R5} )

Using eqs (4) as boundary conditions and the gaugino masses, it is now
straight forward to compute the sfermion masses for the first two 
generation at the EW scale. We obtain
\begin{eqnarray}
\MUL^2 &=& m_0^2 + 0.94 \GLUM^2 + 0.74 D + 0.35 M_Z^2 \: cos2\beta \nonumber \\
\MUR^2 &=& m_0^2 + 0.91 \GLUM^2 + 0.24 D + 0.15 M_Z^2 \: cos2\beta \nonumber \\
\MDL^2 &=& m_0^2 + 0.94 \GLUM^2 + 0.74 D - 0.42 M_Z^2 \: cos2\beta \nonumber \\
\MDR^2 &=& m_0^2 + 0.90 \GLUM^2 - 1.72 D - 0.07 M_Z^2 \: cos2\beta \nonumber \\
\MSNU^2 &=& m_0^2 + 0.05 \GLUM^2 - 2.22 D + 0.5 M_Z^2 \: cos2\beta \nonumber \\
\MLL^2 &=& m_0^2 + 0.05 \GLUM^2 - 2.22 D - 0.27 M_Z^2 \: cos2\beta \nonumber \\
\MLR^2 &=& m_0^2 +  0.02 \GLUM^2 + 1.24 D - 0.23 M_Z^2 \: cos2\beta 
\end{eqnarray}
The numbers quoted in these equations are obtained by using the central
value of $\alpha_G$ given above. 

Using $m_0$ (or equivalently the average left squark mass $\MSQ$),
$\GLUM$, tan$\beta$ and $D$ as the free parameters, one can study the
entire sfermion spectrum. We, however, emphasise that {\bf certain broad mass
patterns} rather than very specific choices of the individual
 masses are responsible for interesting phenomenology.
In the following we identify such patterns which yield  \SUSY  signatures 
 quite distinct from the conventional ones, leading to the possibility
of revision of the current mass limits.

\noindent
A)  If $\mathbf  m_B^2 << \GLUM^2$, where $m_B$ generically represent the 
boundary values in eqs.(4) ,the nondegeneracy among the squarks 
at the EW scale is
negligible in this case with all squarks having a common 
mass close to $\GLUM$. The squark sector is, therefore, very similar to the 
one in the conventional scenario. In contrast, the lepton doublet 
$\tilde{l}$
can still be much lighter compared to the conventional prediction for the same
$\GLUM$ if $D > 0 $.

The striking phenomenological consequences of the light slepton scenario
have already been emphasised in the literature \cite{R17,R18,R19}. Of
particular interest is the case with  $\MSNU<\MCH1$, where $\W1$ is the
 lighter chargino \cite{R18,R19}. As has been discussed
already, in this case the $\tilde{\nu}$-s decay invisibly via the channel
$\tilde{\nu} \to \nu \LSP$, where $\LSP$ is the lightest neutralino assumed
to be the lightest sparticle (LSP),
and the $\W1$ decays via the 2-body decay mode $ \W1 \to l^{\pm}\;\; \SNU$.

This  mass pattern can also be accommodated in the conventional scenario.
However, the available parameter space is rather narrow (see the region
bounded by the two solid lines in Fig. 1, the other relevant \SUSY
parameters are chosen  
to be $\mu = - 300.0$GeV and tan$\beta =2$ ). This perhaps is the reason for
this scenario not being enthusiastically considered in most of the
analyses.  Once nonuniversal scalar masses at the intermediate scale are
taken into account, the region expands significantly as is illustrated by
the other lines  in Fig. 1 which are obtained for different choices of
$D$ (see the figure caption for further details). We have refrained from
using choices numerically close to the extreme value $ D = 0.45 {m_0}^2$
which extends the allowed region even more dramatically.  In obtaining
these regions we have taken into account the lower bounds on $\MSELR,
\MSNU$ and $\MCH1$ derived from LEP \cite{R20}.  Thus we find that
relatively light $\SNU$'s look more probable within the basic framework of
\SUGRA $\;\;$ once the theoretical uncertainities in the GUT scale/Planck
scale physics are taken into account.

One of the most important consequences of a light sneutrino is that
the squark-gluino mass limits obtained at the Tevatron 
 from the search in the $jets+ \not {\!\!\! E}_T$ \cite{R21}
 \hskip 7pt channel are dependent on $\MSNU$ \cite{R19}.
ur analyses reveal that this possibility is not confined to a tiny
region of the parameter space. The effects of a light sneutrino
should therefore be considered very seriously especially  when sparticle
mass limits are quoted for $\MSQ \approx \GLUM$.

Moreover, in view of the uncertainties in the magnitudes of $D$, it is not
advisable to compute  $\MSNU$ from any specific formula. Fortunately, the
precise value of $\MSNU$ is unimportant for estimating its effects on the
squark - gluino mass limits. As shown in \cite{R19} these  limits 
are sensetive to $\Delta m =
\MCH1 - \MSNU$. As long as $\Delta m$ is appreciable (say $>$ 20  GeV ),
one obtains conservative limits on $\GLUM$ and $\MSQ$.  For small $\Delta
m$ , on the other hand, the limits become stronger.  Thus the limits from
the $jets + \not {\!\!\! E}_T$ channel play a complimentary role to those
obtained from the dilepton \cite {R22}channel which become weaker for small 
$\Delta m$ \cite{R19}.

\noindent
{\bf B)} Another interesting region of the parameter space is characterised by
{\bf $\MSQ >> \GLUM$}. In this case the sfermion masses at the EW
 scale are dominated by their boundary values at $M_I$. Thus  sizable
$D$-term contributions may indeed lead to significant mass hierarchies
among the squarks. Such heavy squarks are of course beyond the kinematic
limit of the Tevatron. In this scenario \SUSY  signals at Tevatron are
expected to come primarily from gluino pair production with a cross section
not very sensitive to $D$ (see Table I). Yet a significant
mass split between the L and R squarks may indirectly influence the 
signal by drastically modifying  the gluino branching ratio(BR)s compared
to their values in the conventional scenario.

For $D > 0$, the lightest among the squarks turns out to be of the $\DR$
type. For gluino masses accessible to the Tevatron the LSP is usually
a pure Bino which couples favourably to these light  $\DR$s. As a result
BR ($\GL \rightarrow \LSP + X$) increases significantly since it is
mediated by the light $\DR\;$s. In contrast the conventional scenario
corresponding to degenerate squarks ( $D = 0$ ) predicts dominantly
cascade decays of the $\GL$. Thus the nonuniversality among the squark
masses may lead to a $\not {\! p}_T$  
spectrum much harder than the conventional one.

For $D < 0$, on the other hand, $\LSQ$s are likely to be the lightest species
among the squarks which are favourably coupled to the $\W1$s. Thus compared
to the conventional scenario, the cascade decays of the gluino may become 
even more probable and direct decays of the gluino into the LSP may be reduced
to an insignificant level provided the $\DR$s are sufficiently heavy. This
would soften the   $\not {\! p}_T$ spectrum .

In Table I we present different BRs of $\GL$ decays to $\W1, \LSP $
and $\N2$ (the second lightest neutralino) for some
representative choices of $D$. We have chosen $\GLUM = 250$ GeV and the mass
of the heaviest squark ( either $\LSQ$ or $\DR$ ) to be 1 TeV. Through out
this paper the c.m. energy is taken to be 2.0 TeV corresponding to the
proposed upgrade of the Tevatron.
 It is interesting to note that BR ($\GL \rightarrow
\LSP + X$) varies between the two extremes of our  choices: $D = 0.30
m_0^2$ and $ D = - 0.56 m_0^2$ by a factor as large as 5 ! This strongly
suggests that the effects of nonuniversality on the gluino mass limits
for $\MSQ >> \GLUM$ should be reanalysed. Conservative limits are likely
to be obtained in the $jets+\not{\!\!\! E}_T$ channel for $D < 0$, since this 
case corresponds to the smallest 
production cross section (see Table I) and the softest $\not {\! p}_T$        
spectrum (see below).
In the dilepton channel on the other hand conservative limits are likely
to follow for $D > 0$..In either case the sensitivity of the current limits 
to phenomenological parameter $\LMSQ - \MDR$ is worth reexamination.
In Fig 2 we present the   $\not {\! p}_T$ spectrum 
for the above extreme values
of $D$ as well as for $D = 0$ ( the conventional scenario) using a parton
level Monte Carlo. This illustrates the  effects discussed above qualitatively
. It is well known that a strong cut on    $\not {\! p}_T$
is crucial in seperating
the \SUSY  signal from the background. It is therefore clear that 
nonconventional
gluino branching ratios due to effects of GUT scale/Planck scale physics 
may affect the limits on $\GLUM$ for $\MSQ >> \GLUM$. Another important
constraint on the \SUSY  signal is the  requirement of three or four hard 
jets with $p_T > 25$ GeV \cite{R21}. We have checked that this cut makes
the difference between the three distributions in Fig. 2 even more prominent.
However, we do not present the details here since putting the experimental
cuts on the parton jets may not be completely realistic.

\noindent
{\bf C)} The third region of the parameter space which we find
phenomenologically interesting corresponds to {\bf squarks moderately
heavier than gluinos}.  However, both $\MSQ$ and $\GLUM$ are within the
kinematic reach of the Tevatron. In this case the mass hierarchy among the
squarks may not be as large as in case B) , yet $\LMSQ - \MDR$ can be
numerically significant (see Table II). In Fig.1 the region of the
$\MSQ - \GLUM$ plane above the upper dot dashed line corresponds to
$\MSQ - \MDR > 25 $ GeV for $ D > 0.16 m_0^2$, while that above the upper
dashed line correspond to $\MDR - \MSQ > 25 $ GeV for $ D < - 0.16 m_0^2$.
Even such modest mass splittings  turn out to be
sufficient to make the total squark - gluino production cross section
$\sigma_{\SQ \GL}$ appreciably different from the conventional predictions
corresponding to $D = 0$. The difference is basically driven by the
production of gluinos in association with a relatively light squark.  This
is illustrated in Table II where we present $\sigma_{\SQ \GL}$ for
different choices of $D$. We have chosen $\GLUM = 230$ GeV and the mass of
heaviest squark ($\LSQ$ for $D > 0$ and $\DR$ for $D < 0$) = 300 GeV. Since
there are five squark flavours of the $\LSQ$ type, relatively light
$\LSQ$s affect the cross section quite drastically (see Table II ).

Although in this case the branching ratios of squark/gluino decays 
are rather similar to the conventional ones, the size of the \SUSY  signals 
can be quite different due to the above
modification in the  cross sections. It is to be noted
that for an integrated luminosity of 1 $fb^{-1}$, which can be accumulated
after the luminosity upgrade of the Tevatron, the total number of $\SQ$ -
 $\GL$ events ( Table 2) are strongly 
affected by modest nonuniversalities among the
squark masses, which are quite probable in view of the uncertainties in
Planck scale/GUT scale physics.
 
Although our numerical analyses are restricted to scenarios relevant for
\SUSY searches at Tevatron, we have checked that nonuniversality of sfermion
masses may lead to similar significant effects at LHC. For example,
the heavy squarks discussed  in case B above, are accessible to LHC. Therefore 
 nonuniversality in  their  masses may
affect the \SUSY signals not only through the gluino branching ratios but
also through the total squark-gluino production cross section. For extreme
positive values of $D$, mass hierarchies such as $\LMSQ >\GLUM >\MDR$ are
also allowed.  In this case $Br(\GL \to \LSP + X$) becomes 100\% making
the resulting $\not {\! p}_T$ spectrum very hard.

In conclusion we reiterate that possible variations in GUT/Planck scale
physics may significantly alter the \SUSY signals and the existing mass
limites from the Tevatron, expected in the conventional scenario.
We have identified three mass patterns which might lead to signals quite
distinct from the ones predicted conventionally. All these patterns may arise 
naturally in variations of $N=1$ \SUGRA  motivated models with additional
nonuniversal $D$ terms at a high scale ($M_I$ or $M_G$). 
For $\MSQ \simeq \GLUM$, the existing  mass limits on the sparticles
should be reexamined keeping in view the possibility of a relatively light
sneutrino ($\MSNU< \MCH1$). For $\MSQ >>\GLUM$, nonuniversality in the
suark masses may modify the signal by changing the gluino branching ratios
anticipated in the scenario with universal soft breaking masses for the
sfermions. This may lead to revisions of the gluino mass limits.
For squarks moderately heavier than the gluinos but both 
having masses  within the kinematic limit of
Tevatron, even modest nonuniversality among the squark masses may affect
the signal and the existing squark-gluino mass limits by modifying the
total squark-gluino cross sections. 

Although for the purpose of illustration
we have restricted our numerical results mostly  to $SO(10)$ \SUSY  GUT
 with intermediate 
mass scales, many of our conclusions are also valid even if the GUT directly 
breaks down to the MSSM.

\vskip 25pt

\noindent {\bf Acknowledgement:}
The work of AD was supported by grants from the Department of Atomic Energy
and the Department of Science and Technology, Govt. of India. The work of
Aseshkrishna Datta was supported by the Council of Scientific and Industrial
Research, Govt. of India. MKP thanks Professors G.Senjanovic,A.Yu.Smirnov,
S.Randjbar-Daemi , High Energy Group and the International Centre for Theoretical
Physics, Trieste, Italy for hospitality as ICTP Associate.

\newpage

\noindent
{\large \bf Table Captions :} \\
\vskip 10pt
\noindent
{\bf Table-I :} Branching Ratios for gluino decays (see text) and gluino pair
production cross section for $\MSQ >> \GLUM$ in a $SO(10)$ \SUSY GUT with an
intermediate scale for different values of $D$.

\vskip 10pt 

\noindent
{\bf Table-II :} The total squark-gluino production cross section and the
total number of squark-gluino events for an integrated luminosity of 
1 $fb^{-1}$ with squarks moderately heavier than the gluino in a $SO(10)$ 
\SUSY GUT with an intermediate scale for different values of $D$.

\vskip 40pt
\noindent
{\large \bf Figure Captions :}\\
\vskip 10pt

\noindent 
{\bf Fig. 1} The regions in the $\MSQ$ (average left squark mass)--$\GLUM$
plane where (a) light $\SNU$'s ($\MCH1 >\MSNU$) or (b) significant mass
difference between the L and R squarks may arise in different \SUSY GUTs.
The region bounded by the solid curves satisfies criterion (a) in the 
conventional scenario. The long dashed line represents the upper edge of 
the allowed region (type a) in a grand desert type $SO(10)$ \SUSY GUT with
$D=0.16 {m_0}^2$. The dotted line (lower dot-dashed line) corresponds to the
same upper edge in a $SO(10)$ \SUSY GUTs with an intermediate scale for $D=0$
($D=0.16 {m_0}^2$). In the region above the upper dot-dashed line (short
dashed line) $\LMSQ-\MDR >25$ GeV ($\MDR-\LMSQ >25$ GeV) in an intermediate 
scale $SO(10)$ \SUSY GUT with $D=0.16 {m_o}^2$ ($D= -0.16 {m_o}^2$).
Through out this paper we take $\mu=$ -300 and $tan \beta=$ 2 .

\vskip 10pt 

\noindent
{\bf Fig. 2} The   $\not {\!\! p}_T$ spectrum 
for gluino decays
in $SO(10)$ \SUSY GUT
with an intermediate scale. We have taken $\GLUM=250$ GeV and the heaviest 
squark mass equal to 1 TeV. The solid, dashed and dot-dashed curves
correspond to $D=0$, $D=0.30 {m_0}^2$ and $D= -0.56 {m_o}^2$ respectively.

\newpage

\vskip 25pt
\centerline{Table I}
\vskip 25pt

\begin{center}
\begin{tabular}{|c|c|c|c|c|c|c|c|}
\hline
$D \over {m_o}^2$ & $\LMSQ$ & $\MUR$ & $\MDR$ & $Br(\W1)$ & $Br(\LSP)$
& $Br(\N2)$ & $\sigma_{total}$ in pb\\
\hline
0.30 & 1000 & 930 & 583 & 33.6 & 45.3 & 21.0 & 1.64\\
\hline
0.16 & 1000 & 965 & 817 & 45.5 & 26.4 & 28.1 & 1.67\\
\hline
0 & 1000 & 1000 & 1000 & 49.3 & 20.3 & 30.4 & 1.70\\
\hline
-0.16 & 842 & 876 & 1000 & 51.6 & 16.6 & 31.8 & 1.55\\
\hline
-0.56 & 542 & 661 & 1000 & 56.1 & 9.4 & 34.5 & 1.16\\
\hline
\end{tabular}
\end{center}

\vskip 35pt
\centerline{Table II}
\vskip 25pt
\begin{center}
\begin{tabular}{|c|c|c|c|c|c|}
\hline
$D \over {m_o}^2$ & $\LMSQ$ & $\MUR$ & $\MDR$ & $\sigma_{total}$ in pb &
 No. of events\\
\hline
0.30 & 300 & 288 & 246 & 3.57  & 3570 \\
\hline
0.16 & 300 & 293 & 274 & 3.29 & 3290\\
\hline
0 & 300 & 300 & 300 & 2.99 & 2990\\
\hline
-0.25 & 269 & 274 & 300 & 4.24 & 4240\\
\hline
-0.56 & 242 & 253 & 300 & 6.00 & 6000\\
\hline
\end{tabular}
\end{center}

\begin{figure}[hbt]
\centering
\epsfxsize =15cm
\epsfclipon
\leavevmode
\epsffile{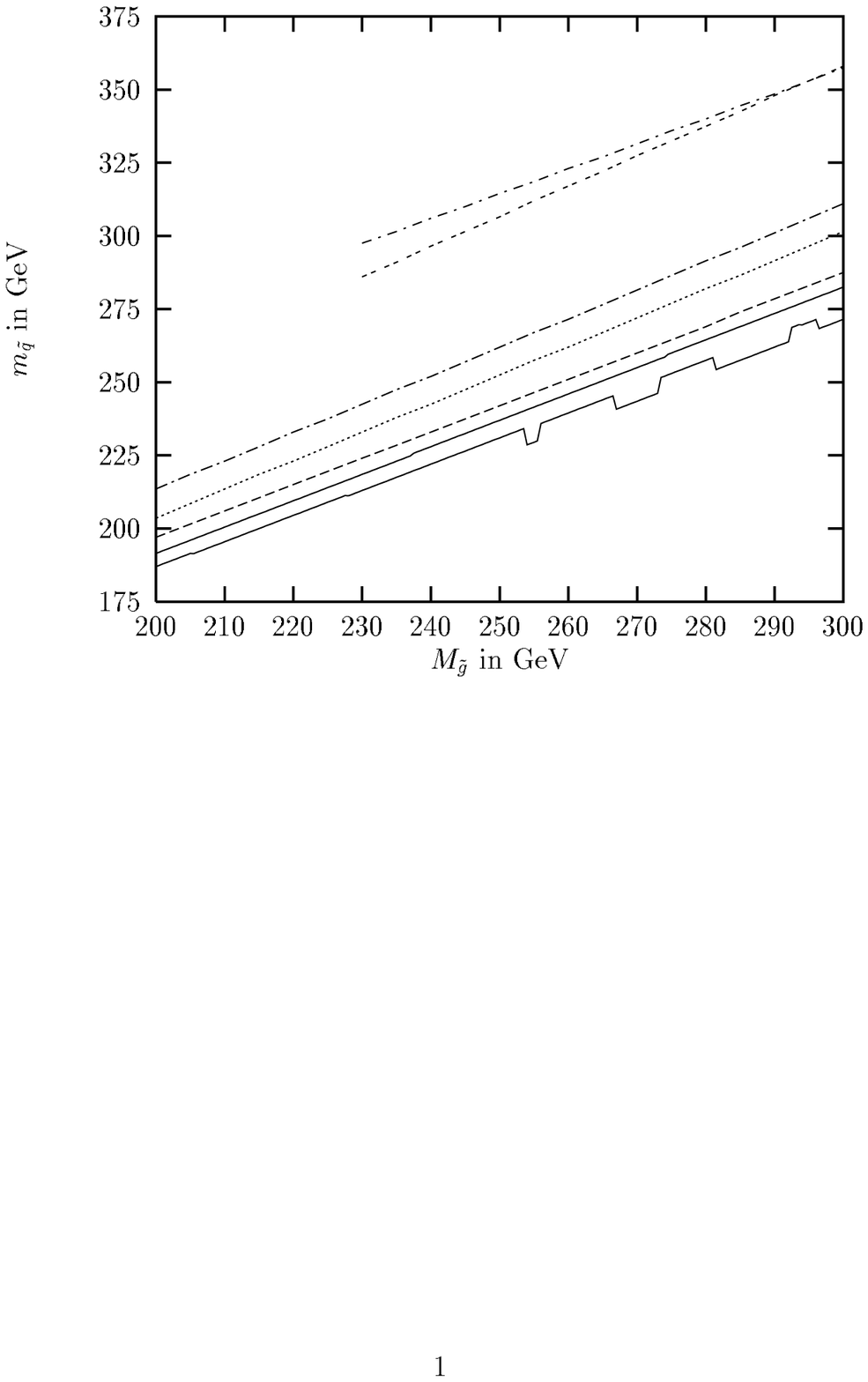}
\vspace{-2.5in}
\caption{}
\label{fig1}  
\end{figure}

\begin{figure}[hbt]
\centering
\epsfxsize =15cm
\epsfclipon
\leavevmode
\epsffile{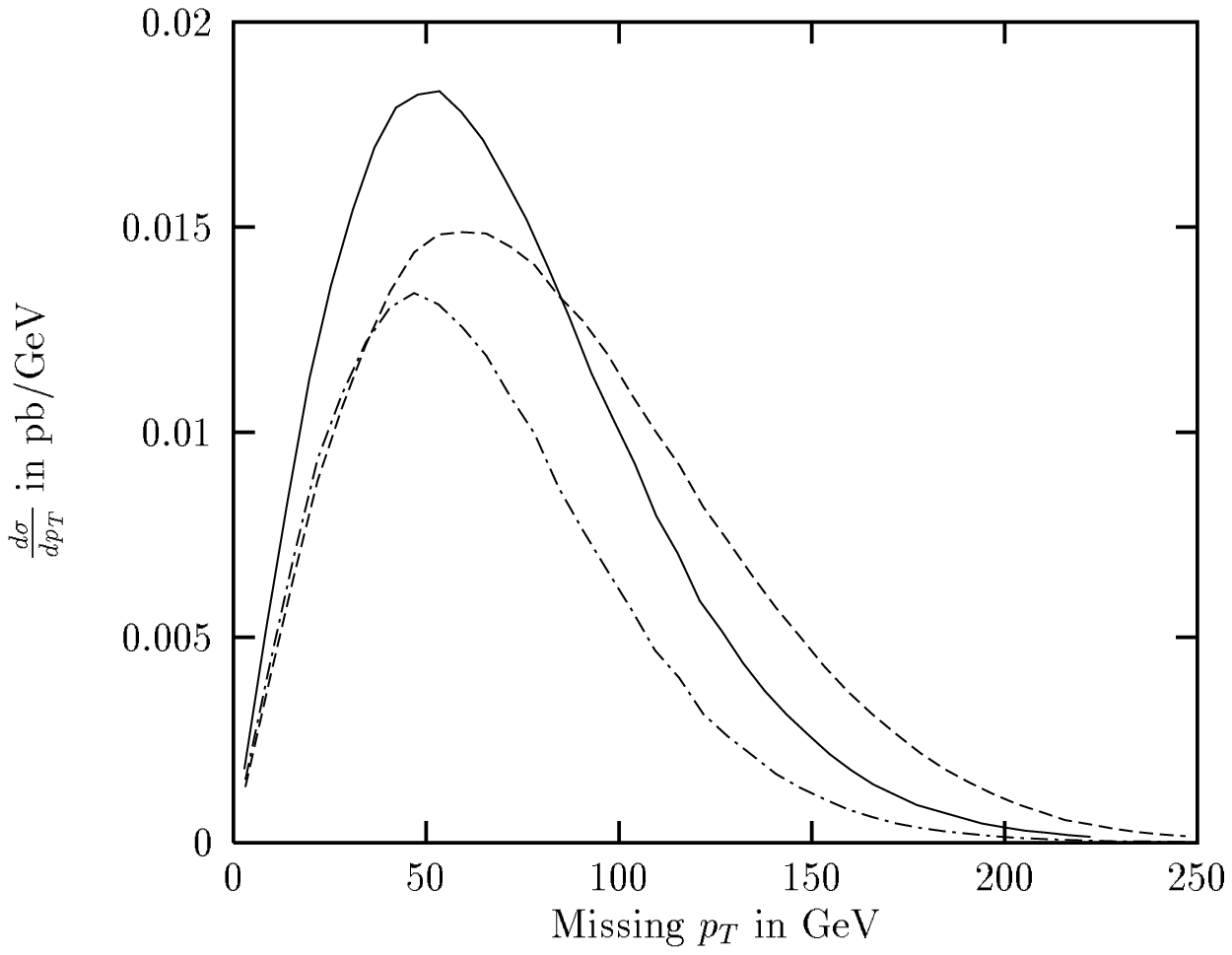}
\vspace{-2.5in}
\caption{}
\label{fig2}  
\end{figure}

\end{document}